\documentclass[prl,aps,preprint,onecolumn,citeautoscript,groupedaddress,noeprint,nofootinbib]{revtex4-2}

\usepackage{amsmath}
\usepackage{bm}
\usepackage{graphicx,color}
\usepackage{subfigure}
\usepackage{array}
\usepackage{braket}
\usepackage{nicefrac}
\usepackage[dvipsnames]{xcolor}
\usepackage{placeins}

\usepackage{color}
\definecolor{myblue}{rgb}{0,0,1}
\usepackage[breaklinks=true,colorlinks=true,linkcolor=myblue,urlcolor=myblue,citecolor=myblue]{hyperref}

\usepackage{verbatim}

\newcommand*{\gw}{{\textit{G}\textsubscript{0}\textit{W}\textsubscript{0}}}
\newcommand*{\GW}{\textit{GW}}

\begin{document}

\title{Unified Deep Learning Framework for Many-Body Quantum Chemistry via Green's Functions}

\author{Christian Venturella}

\author{Jiachen Li}

\author{Christopher Hillenbrand}

\author{Ximena Leyva Peralta}

\author{Jessica Liu}

\author{Tianyu Zhu}
\thanks{tianyu.zhu@yale.edu}

\affiliation{Department of Chemistry, Yale University, New Haven, CT, USA 06520}

\begin{abstract}
Quantum many-body methods provide a systematic route to computing electronic properties of molecules and materials, but high computational costs restrict their use in large-scale applications. Due to the complexity in many-electron wavefunctions, machine learning models capable of capturing fundamental many-body physics remain limited. Here, we present a deep learning framework targeting the many-body Green’s function, which unifies predictions of electronic properties in ground and excited states, while offering physical insights into many-electron correlation effects. By learning the \GW~or coupled-cluster self-energy from mean-field features, our graph neural network achieves competitive performance in predicting one- and two-particle excitations and quantities derivable from one-particle density matrix. We demonstrate its high data efficiency and good transferability across chemical species, system sizes, molecular conformations, and correlation strengths in bond breaking, through multiple molecular and nanomaterial benchmarks. This work opens up opportunities for utilizing machine learning to solve many-electron problems.
\end{abstract}

\maketitle

\newpage

\section{Introduction}
Predicting electronic properties of molecules and materials in ground and excited states is a central task in quantum chemistry and computational materials science. Density functional theory (DFT) has been the primary tool for this task due to balanced accuracy and efficiency~\cite{kohnSelfConsistentEquationsIncluding1965}, but it has well-known systematic errors and uncertainties stemming from approximate exchange-correlation functionals~\cite{cohenChallengesDensityFunctional2012}, which limit its predictive capability. Ab initio many-body electronic structure methods, such as coupled-cluster (CC) theory~\cite{Bartlett2007b} and many-body perturbation theory (\GW)~\cite{Hybertsen1986,Golze_2019}, offer a promising route to more robust quantum mechanical simulations. These methods are particularly desired in the simulations of catalysis and materials that require explicit treatment of electron correlation, such as bond-breaking and excited-state phenomena as well as transition metal compounds. However, their high computational costs prohibit their application to the study of large systems or screening of many molecules.

Data-driven machine learning (ML) has been extensively explored to accelerate quantum chemistry calculations at different levels of theory
~\cite{Keith2021,Westermayr2021b,Deringer2021,VonLilienfeld2020a,Behler2007,Zhang2018c,Schutt2018a,Smith2019,Qiao2020}. These ML models mostly focus on predicting the potential energy or one electronic property (e.g., dipole moment, orbital energy) at a time. Recently, ML models aiming at more fundamental quantum mechanical quantities, such as the mean-field Hamiltonian~\cite{Schutt2019a,Westermayr2021,Li2022deep,Gong2023general}, electron density~\cite{Grisafi2019b,Brockherde2017,ChenghanLi2024c}, Kohn-Sham eigenstates~\cite{Knøsgaard2022representing,Hou2024unsupervised}, and one-particle density matrix~\cite{Shao2023}, start to appear, where various electronic properties can be derived following a single ML prediction. Nevertheless, these methods are usually developed for DFT and limited by its inherent errors, while ML approaches capable of predicting both ground- and excited-state many-body properties within a unified framework remain rare. The main reason is that the size of many-electron wavefunction grows rapidly 
with respect to the molecular size, resulting in patterns that are too complex to learn. Electron density 
computed at the many-body level could serve as the ML target, but directly mapping ground-state electron density information to excited states is a non-trivial task~\cite{Bai2022}. Furthermore, generating many-body quantum chemistry training data is very expensive, which requires the ML method to be highly data-efficient.

In this work, we propose to use the many-body Green's function (MBGF) as the central quantity to enable a deep learning framework that seamlessly connects ground- and excited-state predictions at quantum many-body level. The Green's function $G(\omega)$ is a frequency-dependent quantity that describes the propagation of an electron/hole in a many-electron system. The size of MBGF grows quadratically with respect to the system size, making it a more compact representation of many-body physics compared to the wavefunction. The Green's function theory provides a rigorous road map towards solving the Schr\"{o}dinger equation exactly by simulating the one-particle (charged) and two-particle (neutral) excitations through Hedin's equation and Bethe-Salpeter equation (BSE)~\cite{Hedin_1965,Blase_2020}. In addition, 
MBGF also contains most of the essential ground-state information. Its static limit yields the one-particle density matrix, while integrating MBGF along the imaginary frequency axis gives the ground-state energy. In recent years, ab initio MBGF methods have achieved great success for simulating correlated molecules and materials, based on \GW~\cite{Golze_2019,Hybertsen1986,Zhu2021a,Lei2022}, CC~\cite{Nooijen_1993,Peng2018a,Zhu2019,Laughon2022}, second-order perturbation theory~\cite{Phillips2014,Hirata2015a}, algebraic diagrammatic construction~\cite{Banerjee2023}, density matrix renormalization group~\cite{Ronca2017a}, and quantum Monte Carlo~\cite{Gull2011}. MBGF is also the central quantity in quantum embedding methods including dynamical mean-field theory~\cite{Kotliar2006,Zhu2020,Zhu2021c} and self-energy embedding theory~\cite{Lan2015}. Thus, an MBGF-based ML approach will not only unify predictions of many electronic properties of interest, but also offer fundamental insights into electron correlation effects across a large number of molecular and material problems. 

A major challenge in developing this ML method is to represent the frequency-dependent MBGF matrix 
in a compact and equivariant form, while capturing both local and non-local electron correlations encoded in MBGF. We achieve this by developing a graph neural network (GNN) that directly learns the many-body dynamical correlation potential (i.e., self-energy) on a compact imaginary frequency grid, using orbital-based mean-field features in a symmetry- and polarization-adapted basis. This method enhances our recent work~\cite{venturella2023machine}, offering substantially better capability and accuracy than related works~\cite{Arsenault2014,Dong2024} (see comparisons in Supplementary Section 2), and we name the resulting model MBGF-Net. 
On a series of molecular and nanomaterial benchmark problems, we show that MBGF-Net accurately predicts ground- and excited-state properties, including photoemission and optical spectra, quasiparticle energies and renormalizations, as well as quantities derivable from one-particle density matrix, at the levels of \GW~and coupled-cluster singles and doubles (CCSD). We find that MBGF-Net is highly data-efficient, predicting \GW~frontier quasiparticle energies of QM7/QM9 molecules with mean absolute errors under 0.02 eV using a training set of only 2,000 molecules. Furthermore, we demonstrate promising transferability of MBGF-Net across different chemical species, molecular conformations, system sizes, and electron correlation strengths. In particular, the MBGF-Net model trained exclusively on small silicon nanoclusters predicts excitation spectra of silicon nanoclusters up to four times larger with minimal loss of accuracy, thus demonstrating its potential as a unified framework for ML-accelerated many-body quantum chemistry simulations.

\section{Results}
\textbf{MBGF graph neural network}. While geometric deep learning has been widely used in chemical applications~\cite{Batzner2022,Reiser2022,Qiao2020}, the MBGF is a unique ML target that requires careful attention to molecule featurization and GNN architecture. The Green's function matrix in the frequency domain is defined as
\begin{equation}
    G_{ij}(\omega) =  \Braket{\Psi_0 | {a}_i  
    [\omega - (\hat{H}-E) ]^{-1}  {a}^\dag_j | \Psi_0 } +
    \Braket{ \Psi_0 | {a}^\dag_j  
    [\omega + (\hat{H}-E) ]^{-1}  {a}_i | \Psi_0} 
    \label{eq:MBGF}
\end{equation}
where $\omega$ is the frequency (energy), $|\Psi_0\rangle$ is the ground-state wave function, $\hat{H}$ is the Hamiltonian, $E$ is the ground-state energy, and $a_i$ and $a_j^\dag$ are annihilation and creation operators on orbitals $i$ and $j$. Our goal is to design an orbital-based GNN that predicts the MBGF matrix in a given basis set from DFT or Hartree-Fock (HF) solution, thereby bypassing the expensive many-body quantum chemistry calculation. Similar to many ab initio MBGF theories, instead of directly computing $G(\omega)$, MBGF-Net predicts the self-energy, defined through the Dyson's equation
\begin{equation}
    \Sigma(\omega) = G_0^{-1}(\omega) - G^{-1}(\omega).
\end{equation}
Here, the self-energy $\Sigma(\omega)$ captures dynamical (i.e., energy-dependent) many-body correlation effects missed by the mean-field Green's function $G_0(\omega)$. Thus, the self-energy $\Sigma(\omega)$ is a natural physics-informed $\Delta$-ML target, as $G_0(\omega)$ is always pre-calculated in our workflow. 

\begin{figure}[hbt!]
\centering
\includegraphics[width=1.0\textwidth]{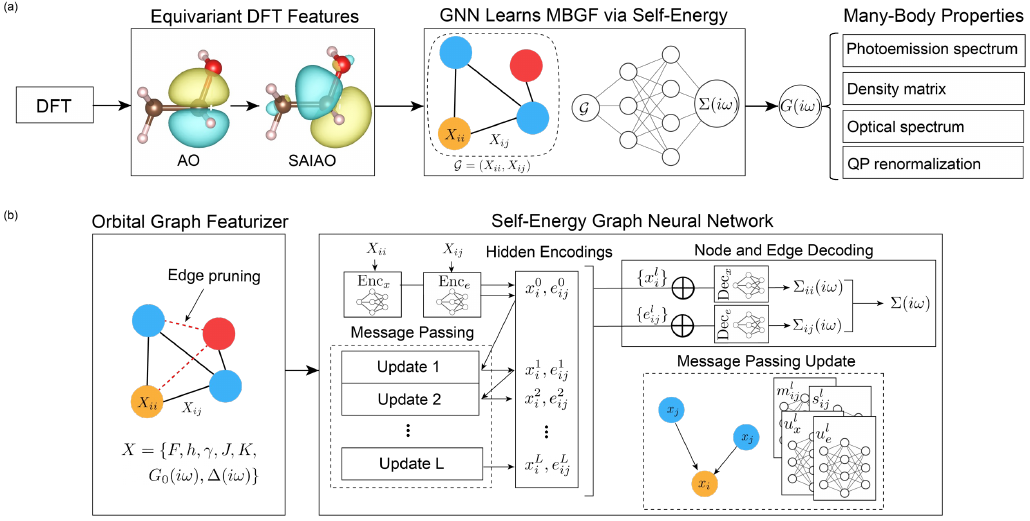}
\caption{Overview of the MBGF-Net workflow and architecture. (a) Starting from a DFT calculation, equivariant DFT features are constructed in the SAIAO basis and mapped onto the orbital graph. After self-energy and MBGF are predicted, MBGF is post-processed to obtain various ground- and excited-state properties at quantum many-body level. (b) The orbital graph is constructed by mapping diagonal and off-diagonal DFT matrix elements to nodes and edges. Edges are pruned based on an orbital interaction criterion. Node and edge features are then autoencoded with corresponding residual blocks. These first hidden encodings are passed to $L$ message passing updates. All $L+1$ hidden node and edge encodings are vector-concatenated ($\bigoplus$) and finally decoded into diagonal and off-diagonal self-energy. 
}
\label{fig:mbgfnet}
\end{figure}

As shown in Fig.~\ref{fig:mbgfnet}a, we adopt an intrinsic atomic orbital plus projected atomic orbital (IAO+PAO) basis~\cite{Knizia2013a} widely used in population analysis and quantum embedding methods to represent all matrices, where the atomic orbitals are polarized by the molecular environment. To ensure rotation invariance, we further apply an angular-momentum block diagonalization step to obtain symmetry-adapted IAO+PAO basis~\cite{Qiao2020,venturella2023machine}, which we refer to as the SAIAO basis. To deal with the continuous frequency dependence in dynamical quantities, we express $\Sigma(i\omega)$ and $G(i\omega)$ on a modified Gauss-Legendre grid along the imaginary frequency axis ($N_\omega = 18 \sim 30$). This choice leads to much smoother self-energy and MBGF for ML, which also allows straightforward post-processing of ML-predicted $G(i\omega)$ to access ground-state and spectral properties. Specifically, by analytically continuing $G(i\omega)$ to real-axis $G(\omega)$, one obtains the photoemission spectrum (i.e., density of states, DOS) with a broadening factor $\eta$
\begin{equation}
\mathrm{DOS}(\omega) = -\frac{1}{\pi} \mathrm{Tr} \left [\mathrm{Im} G(\omega + i\eta) \right ] .
\label{eq:dos}
\end{equation}
Within the \GW~approximation, the optical spectrum can be further computed at a reduced cost by utilizing the ML-predicted quasiparticle (QP) energies via the \GW+BSE formalism~\cite{Blase_2020}. The one-particle reduced density matrix (1-RDM) is obtained through efficient numerical integration
\begin{equation}
    \gamma = \frac{1}{\pi}\int_0^\infty G(i\omega) d\omega .
    \label{eq:1rdm}
\end{equation}
In addition, the self-energy curvature encodes orbital-specific electron correlation strength, indicated by the magnitude of the quasiparticle renormalization (a value between 0 and 1)
\begin{equation}
    Z_i = \left[1 - \frac{\partial \left[\mathrm{Im} \Sigma_{ii}(i\omega)\right]}{\partial \omega} \Big|_{\omega=0} \right]^{-1},
    \label{eq:QP}
\end{equation}
where smaller value of $Z_i$ corresponds to stronger electron correlation in orbital $i$.

In MBGF-Net, we take inspirations from OrbNet~\cite{Qiao2020,Qiao2022} to employ DFT (or HF) electronic matrices as features, while predicting frequency-dependent self-energy vectors for every orbital and orbital pair. 
As shown in Fig.~\ref{fig:mbgfnet}b, DFT matrix elements are mapped onto the nodes and edges of the orbital graph. In addition to static features including Fock ($F$), core Hamiltonian ($h$), Coulomb ($J$), exchange ($K$), and density ($\gamma$) matrices, our own dynamical features (mean-field Green's function $G_0(i\omega)$ and hybridization function $\Delta(i\omega)$) inspired by ab initio MBGF theories~\cite{Golze_2019,Kotliar2006,Zhu2020,Zhu2021c} are also employed, which was found to be more effective for MBGF prediction previously~\cite{venturella2023machine}. To reduce the number of graph edges, 
we prune edges when $\max(|J_{ij}|, |K_{ij}|) < \epsilon$ between an orbital pair $i$ and $j$,
where $\epsilon$ is a small cutoff value. This criterion supposes that two orbitals with negligible bare interaction also have negligible many-body correlation, an assumption similar to integral screening metrics used in low-scaling \GW~techniques~\cite{golze2021}. For edges removed from the orbital graph, the orbital-pair self-energy is set to zero. 

The MBGF-Net architecture is presented in Fig.~\ref{fig:mbgfnet}b, with further technical details provided in the Methods section. We use an encoder-decoder scheme to learn self-energy 
over the nodes and edges of the orbital graph, where the entire frequency response vector is decoded from the orbital hidden encodings. 
Architecture tuning is carried out mainly via two parameters: the number of channels in the message passing layers $N_c$ (that is, the encoding widths) and the number of message passing updates $L$. A larger $N_c$ can improve the capacity of the model to accommodate a more diverse chemical space, while a larger $L$ can express a higher degree of orbital entanglement for stronger correlation or more spatially delocalized electronic structure.

\textbf{Many-body quantum chemical properties in ground and excited states.} We first benchmark the performance of MBGF-Net for predicting various quantum many-body properties in ground and charged excited states on a data set consisting of all QM7 molecules (7,165)~\cite{Rupp2012a} and a subset of QM9 molecules (8,000)~\cite{Ramakrishnan2014a}. We generated the training data at the \gw@PBE0 level in the cc-pVDZ basis set~\cite{Dunning1989,Woon1993} with the PySCF quantum chemistry software package~\cite{Sun2020b,Zhu2021a}, using at most 2,000 molecules for training and reserving the remaining 13,165 molecules for testing. We also augmented the training data with 20 conformers for molecules containing 3 or fewer heavy atoms (660 conformers).
\begin{figure}[hbt!]
\centering
\includegraphics[width=1.0\textwidth]{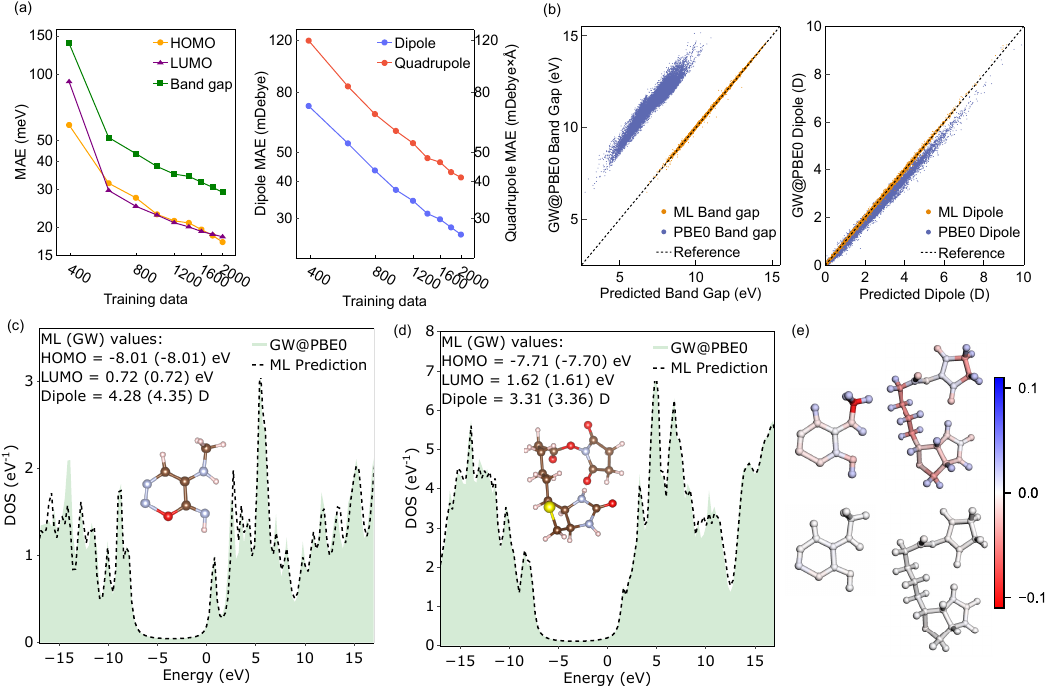}
\caption{MBGF-Net predictions of electronic properties of QM7 and QM9 molecules at the \gw@PBE0 level. (a) Training curves for HOMO, LUMO, band gap, and dipole and quadrupole moments from a single ML model trained using only the self-energy data. MAE stands for mean absolute error. (b) Scatter plots comparing ML-predicted band gaps and dipole moments against PBE0 (baseline) and true \gw@PBE0 values. (c) Prediction of DOS, HOMO, LUMO, and dipole moment on an interpolation case (C\textsubscript{4}H\textsubscript{6}N\textsubscript{4}O). True \gw@PBE0 values are included in the parentheses for comparison. (d) Prediction of same quantities as (c), but on an extrapolation case of NHS-biotin (C\textsubscript{14}H\textsubscript{19}N\textsubscript{3}O\textsubscript{5}S). (e) Comparison of IAO atomic partial charge errors of DFT (top) and ML (bottom) for interpolation and extrapolation cases. Partial charge errors are indicated by the blue or red color.} 
\label{fig:qm9}
\end{figure}

Fig.~\ref{fig:qm9} summarizes the MBGF-Net results, where the baseline DFT calculation used the PBE0 functional~\cite{adamo1999toward} and all quantities were derived from the MBGF-Net model trained exclusively on self-energy data. In Fig.~\ref{fig:qm9}a, five MBGF-Net ensemble models were trained on successively larger training sets. The first subset consisted of the smallest 395 molecules with 5 or fewer heavy atoms, while larger molecules (7 or 9 heavy atoms) were randomly added in the training of subsequent models. We employed a physics-motivated loss function that imposes additional penalties on the self-energy errors on frontier molecular orbitals (FMO) and frequency gradients
\begin{equation}
\mathcal{L} = \mathcal{L}_\mathrm{MSE}(\hat{\Sigma}^\mathrm{SAIAO}, \Sigma^\mathrm{SAIAO}) + \beta_{1}\mathcal{L}_\mathrm{MSE}(\hat{\Sigma}^\mathrm{MO}_{ii}, \Sigma^\mathrm{MO}_{ii} ; i \in \mathrm{FMO}) + \beta_{2}\mathcal{L}_\mathrm{MSE}({\frac{\partial\hat{\Sigma}^\mathrm{SAIAO}}{\partial\omega}}, \frac{\partial\Sigma^\mathrm{SAIAO}}{\partial\omega})
\label{eq:loss}
\end{equation}
where $\hat{\Sigma}$ and $\Sigma$ denote ML-predicted and true self-energy values in SAIAO or molecular orbital (MO) basis, MSE denotes mean-squared error, and $\beta_{1}$ and $\beta_{2}$ control the relative weights of the extra penalty terms. Unless otherwise stated, we set  $\beta_{1}=\beta_{2}=0.1$.

In Fig.~\ref{fig:qm9}a, we find that the mean absolute errors (MAEs) of all ML-predicted quantities drop quickly as more molecules are added into the training set. For the model trained on 2,000 molecules, the MAEs of HOMO (highest occupied molecular orbital) and LUMO (lowest unoccupied molecular orbital) QP energies and band gaps are only 17, 18, and 29 meV, respectively. This performance surpasses that of some state-of-the-art deep learning models (e.g., DimeNet++ and SchNet) on a similar task~\cite{Fediai2023a}. For example, DimeNet++ has larger MAEs for predicting \gw~HOMO, LUMO, and band gap energies (22, 31, 42 meV) on the QM9 data set starting from DFT calculations, even with $10^5$ molecules in the training set ($50\times$ larger than current work)~\cite{Fediai2023a}. From the same model, the 1-RDM predicted by MBGF-Net is of similarly high quality, indicated by small dipole and quadrupole moment MAEs of 26 mD and 41 mD$\cdot$Å. The error distributions are shown in Fig.~\ref{fig:qm9}b, where ML-predicted band gaps and dipole moments are compared against the baseline PBE0 and true \gw@PBE0 values. We then present two case studies in Fig.~\ref{fig:qm9}c,d: a QM9 molecule (C\textsubscript{4}H\textsubscript{6}N\textsubscript{4}O), considered an interpolation task, and a larger molecule, NHS-biotin (C\textsubscript{14}H\textsubscript{19}N\textsubscript{3}O\textsubscript{5}S), considered an extrapolation task. We note that the size of NHS-biotin is more than twice larger than any molecule in the training set. ML-predicted photoemission spectra are in excellent agreement with the \gw@PBE0 spectra for both molecules over a wide energy range. Atomic partial charges derived from the ML-predicted MBGFs also agree perfectly with the true \gw@PBE0 values in both cases (Fig.~\ref{fig:qm9}e). More extrapolation test cases are provided in Supplementary Section 6. Overall, this benchmark demonstrates that MBGF-Net, by learning the many-body electron correlation effects through the self-energy, achieves accurate predictions of many electronic properties with high data efficiency and can generalize to larger molecules well outside the training set.
\begin{figure}[hbt!]
\centering
\includegraphics[width=0.85\textwidth]{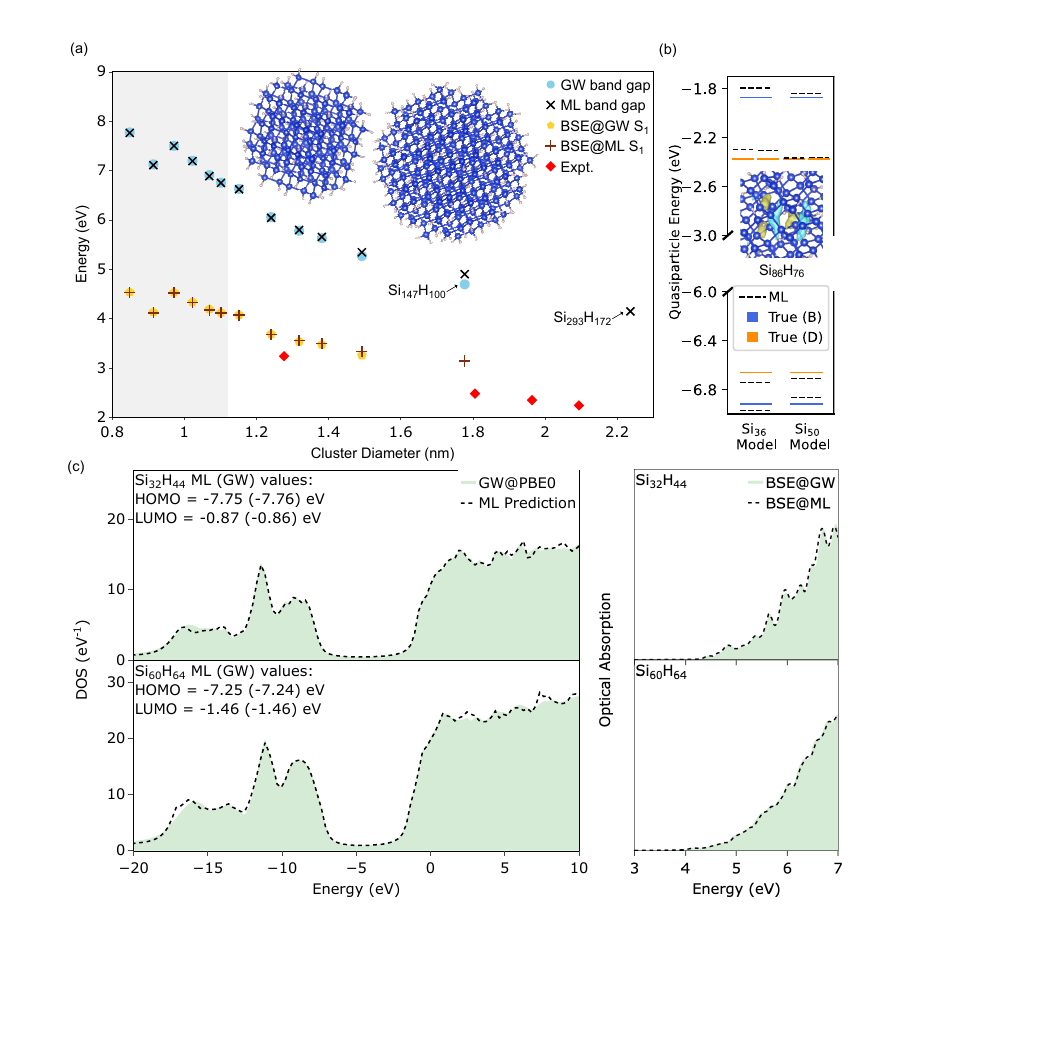}
\caption{MBGF-Net predictions of excited-state properties of silicon nanoclusters at the \gw@PBE0 level. (a) Band gaps and optical gaps (S\textsubscript{1}) for silicon clusters of increasing diameters, assuming a density of 50 Si atoms/nm$^3$ as in bulk silicon. Optical gaps were obtained by solving the BSE equation with true \GW~or ML-predicted QP energies, denoted as BSE@\GW~or BSE@ML. Grey shaded region indicates interpolation regime ($N_\mathrm{Si} \leq 36$ and not seen in the training), while the largest extrapolation cases (structures inset) are Si\textsubscript{147}H\textsubscript{100} (2158 electrons, 6398 orbitals) and Si\textsubscript{293}H\textsubscript{172} (4272 electrons, 12370 orbitals). Experimental optical gap values are taken from Wolkin et al.~\cite{Wolkin1999}. (b) QP energy diagram for predictions of frontier bulk (B) and defect (D) states for Si\textsubscript{86}H\textsubscript{76} with a silicon vacancy using two GNN models trained up to Si\textsubscript{36}H\textsubscript{46} and Si\textsubscript{50}H\textsubscript{56}. The HOMO of the defect state, which is localized at the central vacancy, is inset for reference. The S$_1$ energies from true BSE@$GW$, BSE@Si$_{36}$-Model, and BSE@Si$_{50}$-Model are 2.24, 2.39, and 2.30 eV. (c) Photoemission and optical spectra for Si\textsubscript{32}H\textsubscript{44} (interpolation) and Si\textsubscript{60}H\textsubscript{64} (extrapolation). True \gw@PBE0 values are included in the parentheses for comparison. The lowest 3000 singlet excited states were solved in the \GW+BSE calculations of Si\textsubscript{32}H\textsubscript{44} and Si\textsubscript{60}H\textsubscript{64} using an energy-specifc Davidson algorithm~\cite{hillenbrand2024energyspecificbethesalpeterequationimplementation}
and broadening of 0.05 eV.} 
\label{fig:nanocluster}
\end{figure}
\FloatBarrier
\textbf{Transferability across nanomaterials and molecular systems.} We then demonstrate the transferability of MBGF-Net in more challenging photophysics applications. 
We trained an MBGF-Net model on 159 hydrogenated silicon (Si) nanoclusters with up to 36 Si atoms ($N_\mathrm{Si} \le 36$), for predicting photophysical properties of nanoclusters of sizes up to 293 Si atoms. The training and testing data were generated at the \gw@PBE0 level in the cc-pVTZ basis set, on structures taken from previous works~\cite{zauchner_accelerating_2023, Gao2024}.
In addition to charged excitations, we also utilized the ML-derived \GW~QP energies across the full energy range for the downstream task of computing neutral (optical) excitation energies and spectra via the \GW+BSE formalism (see the Methods section for details). MBGF-Net allows us to bypass the \GW~step in the \GW+BSE calculation, which is more expensive than the BSE step, thus substantially reducing the computational cost. 

As shown in Fig.~\ref{fig:nanocluster}a, MBGF-Net yields near-perfect predictions of band and optical gaps with errors under 25 and 24 meV, not only for Si nanoclusters in the interpolation regime (i.e., $N_\mathrm{Si} \le 36$ and not seen in the training), but also for clusters up to $\sim 2\times$ larger than any training sample (e.g., $\mathrm{Si}_{69}\mathrm{H}_{68}$). This performance outperforms a recently-proposed ML method in accelerating \GW~calculations on the same data set~\cite{zauchner_accelerating_2023}, which suggests our GNN design is effective in capturing long-range screening effects. It should be emphasized that only 11 out of 159 training samples have $N_\mathrm{Si} > 20$, highlighting our method's data efficiency. Even for two large testing clusters up to 4$\times$ larger than any training sample, Si$_{87}$H$_{76}$ and Si$_{147}$H$_{100}$, the band gap errors remain small (75 and 206 meV), which can be further reduced to 6 and 92 meV by adding an additional Si$_{50}$H$_{56}$ cluster into the training set (see Supporting Information Section 7). In Fig.~\ref{fig:nanocluster}b, we show that MBGF-Net predicts accurate defect and bulk QP energies as well as S$_1$ excitation energy for the Si$_{86}$H$_{76}$ cluster with a silicon vacancy (e.g., errors less than 0.06 eV with Si$_{50}$-trained model), although the training data contain only small and non-defective clusters. Fig.~\ref{fig:nanocluster}c shows that MBGF-Net also predicts highly accurate photoemission and optical spectra for Si$_{32}$H$_{44}$ and Si$_{60}$H$_{64}$ over a wide energy range. We note that, while extrapolating to extremely large cluster sizes with similarly high accuracy as interpolation cases remains challenging, it is encouraging that the current MBGF-Net model already predicts errors smaller than other error sources in a standard $G_0W_0$ calculation (e.g., basis set error and DFT dependence) and can be improved with better training data. These results demonstrate impressive transferability of MBGF-Net, which has the potential to enable simulating excited states of large-scale materials beyond the reach of traditional quantum many-body methods.
\begin{figure}[hbt!]
\centering
\includegraphics[width=1.0\textwidth]{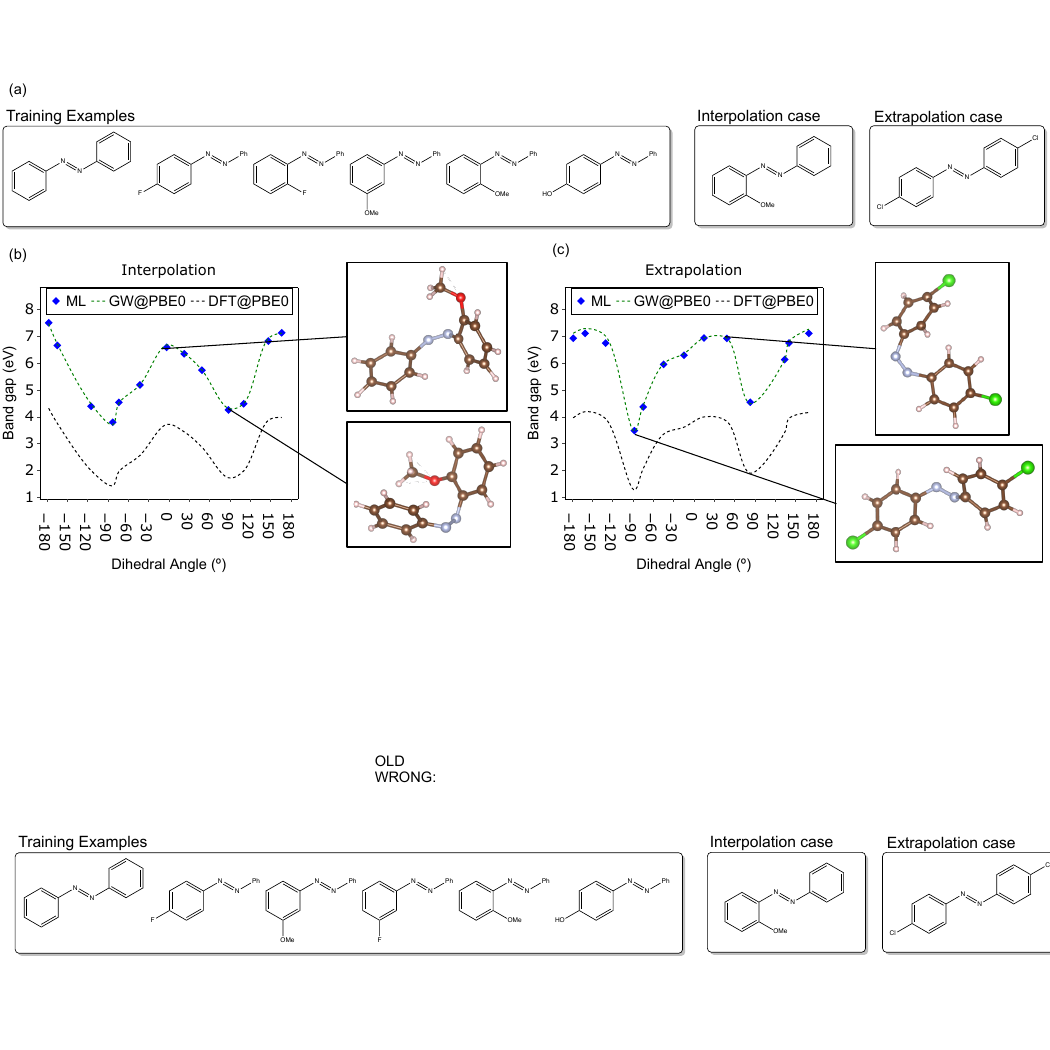}
\caption{MBGF-Net predictions of band gaps of azobenzene derivatives at the \gw@PBE0 level. (a) Summary of training data and test cases. (b) Prediction of band gaps as the CN=NC dihedral angle varies for 2-methoxyazobenzene (interpolation case), compared against true \gw@PBE0 and DFT-PBE0 values. The energy profiles are splined to ease viewing and molecular structures are shown for dihedral angles of $-3^\circ$ and $88^\circ$. (c) Prediction of band gaps as the CN=NC dihedral angle varies for 4,4'-dichloroazobenzene (extrapolation case). Molecular structures are shown for dihedral angles of $-88^\circ$ and $49^\circ$.} 
\label{fig:azobenzene}
\end{figure}
\FloatBarrier
In Fig.~\ref{fig:azobenzene}, we also test whether MBGF-Net can capture subtle electronic structure changes due to conformational distortions, where a model was trained on 100 conformations each of 6 azobenzene derivatives (Fig.~\ref{fig:azobenzene}a), at the \gw@PBE0 level in the cc-pVTZ basis. The full range of the CN=NC dihedral for each derivative was sampled with ab initio molecular dynamics (AIMD) by applying a small bias potential to this torsion using the CP2K software package~\cite{Kuhne2020}. In Fig.~\ref{fig:azobenzene}b, we show that MBGF-Net predicts highly accurate band gaps of 2-methoxyazobenzene conformers with distorted CN=NC dihedral angle, considered an interpolation case. 
Beyond this task, we also applied the same model to 4,4'-dichloroazobenzene (Fig.~\ref{fig:azobenzene}c), which 
falls outside the training data both in terms of atomic composition (chlorine) and substitution pattern (all training examples are singly substituted). ML-predicted band gaps again agree well with true \gw@PBE0 values, suggesting good transferability across conformations and chemical species. To emphasize that MBGF-Net is not simply learning a constant band gap shift, we also provide the energy profiles aligned at minimum band-gap geometry in Supplementary Fig.~13, which show the relative many-body corrections captured by MBGF-Net to be on the order of 0.5 eV.

\textbf{Strong electron correlation in bond-breaking molecules.} We lastly explore the capability of MBGF-Net in the strong electron correlation regime, where quantum many-body treatment beyond DFT (and even \GW) must be used for reliable simulations. We target the cases of C-O single-bond breaking in methanol and ethanol, where CCSD is a reasonable compromise between accuracy and efficiency. We trained an MBGF-Net model on 100 methanol geometries sampled along the C-O stretch and 50 ethane molecules sampled along the C-C stretch (bond length range of 1.3$\sim$4.0 \AA), where the training data were the self-energies computed at the equation-of-motion CCSD level (also known as coupled-cluster Green's function, CCGF)~\cite{Zhu2019,Laughon2022} in the cc-pVTZ basis. Stretched structures were sampled using CP2K metadynamics. Mean-field features were generated at the HF level and the extra loss penalties in Eq.~\ref{eq:loss} were removed in training ($\beta_1=\beta_2=0$).
\begin{figure}[hbt!]
\centering
\includegraphics[width=0.85\textwidth]{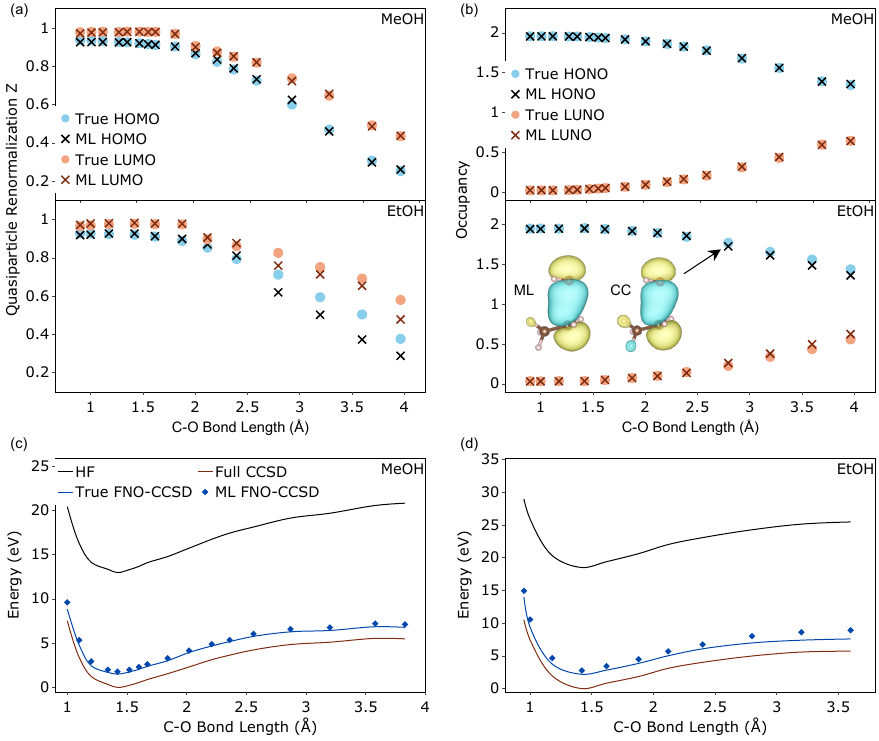}
\caption{MBGF-Net predictions of orbital-specific many-body properties and downstream simulations in C-O single-bond breaking. Training data includes 100 methanol and 50 ethane molecules. (a) Quasiparticle renormalization weights $Z$ of frontier MOs as the C-O bond length increases in methanol and ethanol. (b) Natural occupancies of HONO (highest occupied natural orbital) and LUNO (lowest unoccupied natural orbital) as the C-O bond length increases in methanol and ethanol. ML and true HONOs are shown for ethanol at C-O bond length of 2.8 Å. (c) Ground-state energies of methanol calculated by FNO-CCSD using ML-predicted virtual natural orbitals (threshold $5\times10^{-4}$), compared against HF, FNO-CCSD with true CCSD NOs, and full CCSD. The potential energy curves are not perfectly smooth as the geometries are relaxed using metadynamics. All curves are shifted by the equilibrium CCSD energy. (d) Same as (c), but for ethanol.} 
\label{fig:ccgf}
\end{figure} 
\FloatBarrier
We first show how MBGF-Net uncovers fundamental insights into orbital-specific electron correlation, which cannot be obtained from DFT calculations. QP renormalizations predicted by MBGF-Net are presented in Fig.~\ref{fig:ccgf}a, a quantity commonly used for indicating strength of electron correlation in many-body physics~\cite{Kotliar2006,Zhu2024Kondo}. We find that MBGF-Net achieves near-perfect agreement for HOMO and LUMO $Z$ values of methanol along the C-O stretch, even in the very strongly correlated regime (indicated by small $Z = 0.2 \sim 0.3$). For extrapolating to the ethanol C-O stretch, MBGF-Net exhibits good, but worsened agreement, with systematic errors towards over-correlated $Z$ for a stretched C-O bond, a result of only seeing methanol and ethane in training. In Fig.~\ref{fig:ccgf}b, we derive natural orbitals by diagonalizing the correlated 1-RDM predicted by MBGF-Net: $\gamma V = V n $, where $V$ and $n$ are the natural orbitals (NOs) and NO occupancies. The concept of natural orbitals is widely used in quantum chemistry for analyzing chemical bonding and electron correlation as well as accelerating correlated calculations. MBGF-Net yields accurate predictions of NO occupancies for HONO (highest occupied natural orbital) and LUNO (lowest unoccupied natural orbital) of both methanol and ethanol, where the extent of deviation from integer fillings (0 and 2) indicates the correlation strength along the C-O stretch.


We note that, although it is possible to obtain ground-state energy directly from the integration of MBGF, we leave it to future work due to the numerical sensitivity of this integration to MBGF errors. Instead, we employ ML-predicted CCSD natural orbitals in the downstream task of computing ground-state energies using the frozen natural orbital CCSD (FNO-CCSD) approach~\cite{Taube2008}. By freezing the virtual NOs with natural occupancies smaller than a given threshold ($5\times10^{-4}$ in this work), FNO-CCSD captures a large fraction of CCSD correlation energy at a reduced cost compared to full CCSD, as shown in Fig.~\ref{fig:ccgf}c,d. In this study, the FNO-CCSD virtual space is only 1/3 of the full virtual space, 
corresponding to $\sim$81-fold reduction of computational cost. We find that FNO-CCSD energies based on ML-predicted NOs agree well with those based on true CCSD NOs in the case of methanol, while the agreement deteriorates slightly for ethanol. 
Nevertheless, in both cases, ML FNO-CCSD predictions are substantially better than HF and have small non-parallelity errors compared to full CCSD (errors in energy differences between equilibrium and stretched geometries, see Supplementary Fig.~14). In summary, this benchmark demonstrates that MBGF-Net can serve as a tool for studying correlated electron systems with deep physical insights. In particular, stronger 
many-body electron correlations manifest as larger 
self-energy elements, smaller values of quasiparticle renormalization, and more fractional natural orbital occupancies, which are all obtainable from an MBGF-Net prediction.


\section{Discussion} 

We have developed a deep learning method for predicting quantum many-body properties by targeting MBGF from the DFT electronic structure. We have demonstrated that MBGF-Net achieves near-perfect accuracy on systems similar to the training data, while being transferable to systems well outside the training data, both in terms of size and composition. Our method is also data efficient, needing only hundreds of training molecules in most benchmarks, which ameliorates the high costs of generating many-body quantum chemistry data. 
Even in challenging extrapolation cases that expose limitations of our method, we argue that useful insights into the nature of electronic interactions can still be drawn. For example, the overestimation of band gaps in Si nanoclusters (e.g., Si\textsubscript{147}H\textsubscript{100}) much larger than training samples suggests the long-range nature and length scale of screened Coulomb interactions. While the current work is limited to molecular systems, MBGF-Net can be extended to solids, for example by deriving a symmetry-adapted basis with periodic boundary conditions. MBGF-Net introduces a high-level, many-body capability to the powerful toolbox of Hamiltonian machine learning. If integrated with deep learning approaches for DFT Hamiltonians, it could enable the study of electron correlation effects in systems much larger than those demonstrated here. Moreover, MBGF-Net can be seamlessly integrated into widely-used ab initio MBGF frameworks, e.g., as data-driven impurity solvers within Green's function embedding methods~\cite{Zhu2020,Zhu2021c,Li2024ibdet}, for simulating correlated electron materials. Overall, this work establishes a unified ML framework for many-body quantum chemistry and demonstrates the possibility of ML-accelerated computational study of many-electron systems towards quantitative accuracy.

\section{Methods}

\textbf{Intrinsic atomic orbital plus projected atomic orbital basis.}
Here we briefly review the intrinsic atomic orbital plus projected atomic orbital (IAO+PAO) basis~\cite{Knizia2013a,cui2019efficient}.
With a converged SCF calculation,
the mean-field wave function $|\Psi \rangle$ is defined by the occupied molecular orbitals (MOs) $\ket{\psi_i}$:
\begin{equation}
    |\psi_i \rangle = \sum_{\mu \in B_1} C_{\mu i} | \phi_{\mu} \rangle \text{,}
\end{equation}
where atomic orbitals (AOs) $| \phi_{\mu} \rangle$ are in a large set $B_1$.
To provide a direct connection between quantum chemistry and empirical chemical concepts,
free-atom AOs of the minimal basis $B_2$ are constructed to expand the wave function $|\Psi \rangle$. 

To construct IAOs,
the free-atom AOs 
in $B_2$ is split into a depolarized occupied space $\tilde{O} = \sum_{\tilde{i}} | \psi_{\tilde{i}} \rangle \langle \psi_{\tilde{i}} |$ ($\ket{\psi_{\tilde{i}}}$ are the depolarized MOs defined below) and its complement $\text{I} - \tilde{O}$.
We define the projectors onto the $B_1$ and $B_2$ basis sets as
\begin{align}
    P_{12} = & \sum_{\mu \nu \in B_1} |\phi_{\mu} \rangle S_{\mu \nu} \langle \phi_\nu| \\
    P_{21} = & \sum_{\sigma \tau \in B_2} |\phi_{\sigma} \rangle S_{\sigma \tau} \langle \phi_\tau|
\end{align}
where $S_{\mu \nu}$ and $S_{\sigma \tau}$ are inverse overlap matrices of $B_1$ and $B_2$ basis sets.
The depolarized MOs are obtained from projecting occupied MOs $| \psi_i \rangle$ from the main basis $B_1$ to the minimal basis $B_2$
\begin{equation}
    \{ | \psi_{\bar{i}} \rangle \} = \text{orth} \left (
    P_{12} P_{21} | \psi_i \rangle
    \right ) \text{,}
\end{equation}
where ``orth'' means L\"owdin orthogonalization.
The polarized AOs $| \phi_{\rho}^\mathrm{IAO} \rangle$ can then be obtained from projecting contributions of the free-atom AOs $| \psi_{\tilde{\rho}} \rangle$ ($\tilde{\rho} \in B_2$) in $\tilde{O}$ and $\text{I} - \tilde{O}$ onto the polarized counterparts $O = \sum_i | \psi_i \rangle \langle \psi_i |$ and $\text{I} - O$
\begin{equation}
    | \phi_{\rho}^\mathrm{IAO} \rangle = \left [ 
    O\tilde{O} + (\text{I} - O) (\text{I} -\tilde{O})
    \right ] 
    P_{12} | \phi_{\tilde{\rho}} \rangle \text{,}
\end{equation}
which are labled as IAOs. We refer the readers to the original IAO paper~\cite{Knizia2013a} for more details.

After the IAOs are constructed,
the PAOs are obtained by projecting the IAO component out from the AOs in the main basis $B_1$ to represent the remaining virtual space
\begin{equation}
    | \phi_{\mu}^{\mathrm{PAO}} \rangle =  \sum_{\rho} \big(
    \text{I} -
    | \phi_{\rho}^{\mathrm{IAO}} \rangle \langle \phi_{\rho}^{\mathrm{IAO}} |
    \big)
    | \phi_{\mu} \rangle
\end{equation}
The union of IAO and PAO spaces (i.e., IAO+PAO basis) spans the full AO space.


To ensure rotational invariance in our machine learning framework,
the symmetry adaptation procedure is applied to the IAO+PAO basis~\cite{venturella2023machine}.
To achieve this,
the mean-field Fock matrix in the IAO+PAO basis is diagonalized in each matrix block of shell $(n,l)$ on every atom A
\begin{equation}
    F^{\text{A}}_{nl} Y^{\text{A}}_{nl} = Y^{\text{A}}_{nl} \lambda^{\text{A}}_{nl} \text{,}
\end{equation}
where $n$ and $l$ are principal and angular quantum numbers,
$Y^{\text{A}}_{nl}$ and $\lambda^{\text{A}}_{nl}$ are the eigenvectors and eigenvalues of Fock matrix block $F^{\text{A}}_{nl}$.
The resulting $Y^{\text{A}}_{nl}$ from all blocks are combined into an orthogonal and block-diagonal transformation $Y$ to the symmetry-adapted intrinsic atomic orbital plus projected atomic orbitals (SAIAOs)
\begin{equation}
    | \phi^{\text{SAIAO}}_{\mu} \rangle = \sum_{\nu} Y_{\nu \mu} 
    | \phi^{\text{IAO+PAO}}_{\nu} \rangle \text{.}
\end{equation}
A demonstration of the effectiveness of IAO+PAO transformation compared to symmetry adaption without projection can be found in Supplementary Section 3.

\textbf{Orbital features and graph construction.} In addition to static features, we utilize a set of dynamical features devised in our previous work ~\cite{venturella2023machine}, $G_0(i\omega_k)$ and $\Delta(i\omega_k)$, evaluated on a coarse frequency grid: $\omega_k=[10^{-3}, 0.1, 0.2, 0.5, 1.0, 2.0]$ a.u. $G_0(i\omega_k)$ is the mean-field Green's function
\begin{equation}
    \mathbf{G}_0(i\omega_k) = (\epsilon_F + i\omega_k-\mathbf{F})^{-1},
\end{equation}
where $\epsilon_F$ is the fermi level (taken as the midpoint of the DFT HOMO and LUMO energies). The orbital-based hybridization function $\Delta(i\omega_k)$ are defined separately for nodes and edges. For orbitals, i.e., graph nodes, we define $\Delta_{ii}(i\omega_k)$ as
\begin{equation}
    \Delta_{ii} (i\omega_k) = \epsilon_F + i\omega_k - F_{ii} - [G_0(i\omega_k)]_{ii}^{-1}
\end{equation}
and for orbital pairs, i.e., graph edges, we define $\Delta_{ij}(i\omega_k)$ as
\begin{equation}
    \Delta_{ij} (i\omega_k) = \left\{ \epsilon_F + i\omega_k - F_{ij} - [G_0(i\omega_k)]_\mathrm{sub}^{-1} \right\}_{01},
\end{equation}
where $[G_0(i\omega)]_\mathrm{sub}$ is
\begin{equation}
    (G_0)_\mathrm{sub} \equiv \begin{bmatrix}
    (G_0)_{ii} & (G_0)_{ij} \\
    (G_0)_{ji} & (G_0)_{jj} 
    \end{bmatrix}.
\end{equation}
To ensure stable GNN learning, node and edge features are further transformed by simple deterministic functions and data-dependent scaling. Before node features are passed to the GNN, they are mean-shifted and scaled by their standard deviations over all diagonal elements in the training data (i.e., each feature is standardized). Each edge feature undergoes a sign-preserving log transformation. First, off-diagonal elements are split into positive and negative values $\tilde{X}_{ij}$ from the original $X_{ij}$ as:
\begin{equation}
    X_{ij}^{+} = \begin{cases} 
 \log(|X_{ij}|) & \text{if }X_{ij} > 0 \\ 
0 & \text{else} 
\end{cases}, \qquad X_{ij}^{-} = \begin{cases} 
 \log(|X_{ij}|) & \text{if }X_{ij} < 0 \\ 
0 & \text{else} 
\end{cases}.
\end{equation}
Then, the mean and standard deviation of the nonzero elements of $X_{ij}^{+}$ and $X_{ij}^{-}$ are used to standardize $X_{ij}^{+}$ and $X_{ij}^{-}$, respectively, to obtain $\tilde{X}_{ij}^{+}$ and $\tilde{X}_{ij}^{-}$. Then the final transformed $\tilde{X}_{ij}$ is a concatenation of $\tilde{X}_{ij}^{+}$ and $\tilde{X}_{ij}^{-}$: 
\begin{equation}
    \tilde{X}_{ij} = \{ \tilde{X}_{ij}^{+} \oplus \tilde{X}_{ij}^{-}\}.
\end{equation}
Note that this procedure increases the dimension of untransformed edge feature vectors by a factor of 2. In addition to these mean-field features from our previous work, in this work we utilize the spatial extent of each orbital and IAO atomic partial charges on the atom associated with a given orbital. We employ these only on the nodes and consider them static due to lack of any frequency dependence. For node features, we take care to account for the potentially large magnitudes of $h_{ii}$ and $J_{ii}$.  These elements cancel each other out, 
so we use their sum as the feature to avoid numerical problems when unseen (larger than training) systems are passed to MBGF-Net. In addition to continuous-valued features, we also utilize simple binary (i.e., one-hot) features on the nodes, corresponding to orbital type (core/IAO/PAO), principal number ($n$), and angular momentum number ($\ell$). To clarify how all these processed features are mapped onto orbital graphs, a schematic illustration of this process  is given for an example set of static features in Supplementary Figure 1.\par

Since we focus on valence spectral properties, the self-energy elements associated with core orbitals are set to zero. 
During the SAIAO construction step for the Si nanocluster and bond breaking benchmarks, the core orbitals are projected separately, to avoid mixing the SAIAO core orbitals with the valence and virtual orbitals.

\textbf{GNN architecture.} As shown in Fig.~\ref{fig:mbgfnet}b, two residual networks each encode the nodes and edges. Because we suppose the static, dynamical, and binary features carry different types of information, they are initially passed to three independent linear layers before concatenation in the deeper portion of the residual networks. Two independent decoders are used for node and edge self-energy, though the architecture for each is identical. Besides the two key parameters discussed in the main text ($N_c$ and $L$), all other hyperparameters control these encoder residual networks for nodes ($x$) and edges ($e$). Each encoder is comprised mainly of two disconnected linear layers that process different feature input streams in parallel: one of width $n^{x/e}_\mathrm{enc,st}$ for static feature inputs (denoted $1s$) and another  of width $n^{x/e}_\mathrm{enc,dyn}$ that processes the dynamical features ($1d$). The node encoder has a third disconnected linear layer of width $n^x_\mathrm{enc,bin}$ that processes binary features ($1b$).  Denoting the DFT node/orbital static, dynamical, and binary features as $X_{ii}^s$, $X_{ii}^d$, and $X_{ii}^b$ respectively, the residual network for nodes/orbitals is:
\begin{equation}
    h_i = \sigma(\mathbf{W}_2\cdot\{\sigma(\mathbf{W}_{1s}\cdot X_{ii}^s + \mathbf{b}_{1s}) \oplus \sigma(\mathbf{W}_{1d}\cdot X_{ii}^d + \mathbf{b}_{1d}) \oplus \sigma(\mathbf{W}_{1b}\cdot X_{ii}^b + \mathbf{b}_{1b})\} + \mathbf{b}_2)
\end{equation}
\begin{equation}
    x_i^0 = \sigma(\mathbf{W}_3\cdot h_i + \mathbf{W}_r\cdot \{ X_{ii}^s \oplus X_{ii}^d\}+  \mathbf{b}_3)
\end{equation}
where $h_i$ denotes the hidden part of the encoder and $r$ denotes ``residual" (i.e., the skip connection). The bold tensors $\mathbf{W}$ and $\mathbf{b}$ are the trainable weight matrices and trainable bias vectors respectively. The activation function $\sigma$ is the SiLU nonlinearity and $\oplus$ denotes vector concatenation. \par 
After node and edge features are encoded, $x^0_i$ and $e^0_{ij}$ are passed to the first of $L$ message passing updates. For communication between orbitals $i$ and $j$ within each update, the following transformation of their node features $x$ ensures pair permutation invariance without loss of degrees of freedom:
\begin{equation}
  \tilde{x}_i = x_i + x_j, ~~ \tilde{x}_j = |x_i - x_j|.
\end{equation}
Then, single-headed attentional aggregation is employed for each update $l$ with associated message function $m^l$ gated by attention score function $s^l$. Graph convolution is then followed by single layer node update $u_x^l$ and edge update $u_e^l$:
\begin{equation}
x^{l+1}_i = u_x^l\left(\sum_{j\in \mathcal{N}(i)} \frac{\exp(s^l_{ij})m^l_{ij}}{\sum_{j\in \mathcal{N}(i)}{\exp(s^l_{ij}})}\right), ~~ e^{l+1}_{ij} = u_e^l(m^l_{ij}).
\end{equation}  The message function has the following structure:
\begin{equation}
    m^l_{ij} = \sigma(\mathbf{W}^l_2 \cdot\{\sigma(\mathbf{W}^l_{1i} \cdot\tilde{x}^l_i) \oplus\sigma(\mathbf{W}^l_{1j}\cdot \tilde{x}^l_j)\oplus e^l_{ij}\} + \mathbf{b}^l_2)
\end{equation}
 Note that the second linear layer (with $\mathbf{W}^l_2$ and $\mathbf{b}^l_2$) maps $3N_c$ to $N_c$. The score function has a similar structure (with $\mathbf{W}^l_2$ mapping $3N_c$ to $1$):
\begin{equation}
    s_{ij}^l = \sigma_s(\mathbf{W}^l_2 \cdot\{(\mathbf{W}^l_{1i} \cdot\tilde{x}^l_i) \oplus (\mathbf{W}^l_{1j}\cdot \tilde{x}^l_j)\oplus e^l_{ij}\})
\end{equation}
where $\sigma_s$ is the Tanhshrink nonlinearity. The decoders for nodes ($ii$) and edges ($ij$) are identical, but separate, and have the following structure(s):
\begin{equation}
\mathrm{Re}[\Sigma_{ii}(i\omega)] = \mathbf{W}^x_{2,\mathrm{re}}\cdot\sigma(\mathbf{W}^x_{1,\mathrm{re}}\cdot(\bigoplus_{l = 0}^L x^l_i) + \mathbf{b}^x_{1,\mathrm{re}}) + \mathbf{b}^x_{2,\mathrm{re}}
\end{equation}
\begin{equation}
\mathrm{Im}[\Sigma_{ii}(i\omega)] = \mathbf{W}^x_{2,\mathrm{im}}\cdot\sigma(\mathbf{W}^x_{1,\mathrm{im}}\cdot(\bigoplus_{l = 0}^L x^l_i) + \mathbf{b}^x_{1,\mathrm{im}}) + \mathbf{b}^x_{2,\mathrm{im}}
\end{equation}
\begin{equation}
\mathrm{Re}[\Sigma_{ij}(i\omega)] = \mathbf{W}^e_{2,\mathrm{re}}\cdot\sigma(\mathbf{W}^e_{1,\mathrm{re}}\cdot(\bigoplus_{l = 0}^L e^l_{ij}) + \mathbf{b}^e_{1,\mathrm{re}}) + \mathbf{b}^e_{2,\mathrm{re}}
\end{equation}
\begin{equation}
\mathrm{Im}[\Sigma_{ij}(i\omega)] = \mathbf{W}^e_{2,\mathrm{im}}\cdot\sigma(\mathbf{W}^e_{1,\mathrm{im}}\cdot(\bigoplus_{l = 0}^L e^l_{ij}) + \mathbf{b}^e_{1,\mathrm{im}}) + \mathbf{b}^e_{2,\mathrm{im}}
\end{equation}
The hidden layer width of each decoder is equal to the dimension of concatenated hidden encodings (i.e., $n^x_{dec}=n^e_{dec}=(L+1)\cdot N_c$) and thus is not a tunable parameter. All activation functions $\sigma$ are SiLU functions, except for the score activation function $\sigma_s$. Hyperparameters for each learning task are given in Supplementary Table 1, which control DFT graph construction, GNN architecture, extra loss terms, and Adam training. The core data processing and architecture for our GNN are implemented with the PyTorch and Pytorch Geometric libraries~\cite{paszke2017automatic,Fey:2019wv}, with the attentional aggregation implementation from previous works~\cite{veličković2018graph,li2019graph}.

\textbf{Self-energy loss.} In Eq. \ref{eq:loss}, aside from changing values of $\beta_i$, tuning for specific applications is mostly done via the definition of the FMOs. If the set of FMOs is too small (e.g., fitting to only the HOMO/LUMO self-energy), we observe the training loss oscillates unfavorably in the later part of the optimization. For the QM9 molecules, we find including the range of HOMO $-$ 10 to LUMO $+$ 10 is reasonable, but fixing this range to a constant number is problematic for datasets with highly variable system sizes. For example, with silicon nanoclusters, we instead employ a scheme where percentages of occupied/virtual are included into the set of FMOs. We forego any parameter search and select 30\%/25\% for occupied/virtual MOs to capture all valence occupied orbitals and virtual orbitals up to approximately 20 eV.

\textbf{Active learning strategies.} For QM9, we adopt an ensembling approach to obtain more generalizable and uncertainty-aware predictions of self-energy. For an ensemble of $M$ MBGF-Net models, the predicted self-energy tensor is taken as the mean of the $M$ self-energy predictions:
\begin{equation}\label{eq:al_mean}
\Bar{\Sigma}(i\omega) = \frac{1}{M}\sum_m^M \Sigma^m(i\omega)
\end{equation}
Based on previous works~\cite{Gastegger2017}, we calculate self-energy uncertainty tensor $\Sigma_\sigma(i\omega)$ as
\begin{equation}\label{eq:al_uncertainty}
\Sigma_\sigma(i\omega) = \sqrt{\frac{1}{M-1}\sum_m^M (\Bar{\Sigma}(i\omega) - \Sigma^m(i\omega))^2}
\end{equation}
The gives mean and uncertainty in the equivariant SAIAO basis. Error analysis techniques can be employed to obtain uncertainty in the MO basis from an uncertainty tensor in the SAIAO learning basis:
\begin{equation}\label{eq:al_rotation}
(\Sigma^{MO}_\sigma(i\omega))_{ij} = (\Tilde{C}^T\Tilde{\Sigma}_\sigma(i\omega)\Tilde{C})^{1/2}_{ij}
\end{equation}
where $\tilde{X}$ denotes the elementwise square of $X$. We explore the effectiveness of active learning approaches for learning MBGF for the QM7/QM9 learning task to deal with the high diversity of chemical compositions. In particular, we observe that this dataset has relatively few molecules with functional groups of potential interest to photophysical applications – for example, QM7/QM9 have very few sulfur/fluorine containing compounds. This suggests that a naively trained MBGF-Net may have a bias that reflects the inherent imbalance of the underlying data. To address this, we implement a training strategy we call ``active refinement" that optimizes a pre-trained ensemble model with a focus on the most unusual training examples - the detailed algorithm can be found in Supplementary Algorithm 1. For Fig.~\ref{fig:qm9}, we employ this active refinement for each model, but each addition of new training data is still sampled randomly. In a separate study, we use self-energy uncertainty quantification to more efficiently select new training examples from unseen data. In particular, for the last three points of our training curves in Fig.~\ref{fig:qm9} (1600, 1800, 2000 molecules), we employ active updates to the training data and compare the performance to random updates in Fig.~\ref{fig:qm9}. These results are presented in Supplementary Fig.~12, which show a slight improvement of MAEs in predcited band gap by $\sim2$ meV.

\textbf{Computational Details of the Bethe–Salpeter Equation.}
The quasiparticle (QP) energies obtained from $GW$ are used in the Bethe–Salpeter equation (BSE) to calculate optical excitation energies.
With the static approximation for the screened interaction that treats the frequency as zero,
the working equation of BSE is a generalized eigenvalue problem~\cite{krauseImplementationBetheSalpeter2017}
\begin{equation}\label{eq:bse}
    \begin{bmatrix}
        \mathbf{A} & \mathbf{B} \\
        \mathbf{B^*} & \mathbf{A^*}
    \end{bmatrix}
    \begin{bmatrix}
        \mathbf{X} \\
        \mathbf{Y}
    \end{bmatrix}
    = \Omega
    \begin{bmatrix}
        \mathbf{I} & \mathbf{0} \\
        \mathbf{0} & \mathbf{-I}
    \end{bmatrix}
    \begin{bmatrix}
        \mathbf{X} \\
        \mathbf{Y}
    \end{bmatrix}
\end{equation}
where $\Omega$ is the excitation energy.
In the following, we use $i$, $j$, $k$, $l$ for occupied orbitals, 
$a$, $b$, $c$, $d$ for virtual orbitals, 
$p$, $q$, $r$, $s$ for general molecular orbitals.
In Eq.~\ref{eq:bse}, the $\mathbf{A}$, $\mathbf{B}$ matrices are defined as
\begin{align}
    A_{ia,jb} &= \delta_{ij} \delta_{ab} \left (\epsilon_a^{\text{QP}}-\epsilon_i^{\text{QP}} \right ) + v_{ia,jb} - W_{ij,ab} \\
    B_{ia,jb} &= v_{ia,bj} - W_{ib,aj}
\end{align}
where the bare Coulomb interaction is
\begin{equation}
    v_{pq,rs} =
    \int dx_{1}dx_{2}\frac{\psi_{p}^{*}(x_{1})\psi_{r}^{*}(x_{2})\psi_{q}(x_{1})\psi_{s}(x_{2})}{|\mathbf{r_{1}}-\mathbf{r_{2}}|} 
\end{equation}
and the static screened interaction is
\begin{equation}
    W_{pq,rs} = \sum_{tu} \left (D^{-1} \right )_{pq,tu} v_{tu,rs}
\end{equation}
The dielectric function $D$ is calculated from the static response function $\chi$~\cite{krauseImplementationBetheSalpeter2017}
\begin{align}
    D_{pq,rs} &= \delta_{pr}\delta_{qs} - v_{pq,rs}\chi_{rs,rs} \text{.} \\
    \chi_{ia,ia} &= \chi_{ai,ai} = \left (\epsilon_i^{\text{QP}} - \epsilon_a^{\text{QP}} \right )^{-1} \label{eq:response}
\end{align}
The scaling of solving Eq.~\ref{eq:bse} is $\mathcal{O}(N^4)$ by using the canonical Davidson algorithm,
where $N$ is the size of the system. 
For large systems, we solve BSE in an active space to discard high virtual orbitals and core orbitals that have negligible contributions to the desired low-lying excited states. In particular for silicon nanoclusters, 
the BSE equation is solved in the active space that included all valence occupied orbitals and unoccupied orbitals with QP energies under 20 eV. We note that, BSE with ML-predicted QP energies is substantially cheaper than the true \textit{GW}+BSE calculation, since solving active-space BSE takes much less time and memory than a full \textit{GW} calculation. We further refer the readers to Hillenbrand et al.~\cite{hillenbrand2024energyspecificbethesalpeterequationimplementation} for the implementation details.

\section{Data availability}
Source data for Figures~\ref{fig:qm9},~\ref{fig:nanocluster},~\ref{fig:azobenzene},~\ref{fig:ccgf} are provided with this paper. The datasets used in this work are available at Zenodo~\cite{zenodo_data}.

\section{Code availability}
The code used in this work is available at Zenodo~\cite{zenodo_code} and on GitHub (\href{https://github.com/ZhuGroup-Yale/mlgf}{https://github.com/ZhuGroup-Yale/mlgf}). Its implementation uses the fcDMFT code at \href{https://github.com/ZhuGroup-Yale/fcdmft}{https://github.com/ZhuGroup-Yale/fcdmft} and PySCF at \href{https://github.com/pyscf/pyscf}{https://github.com/pyscf/pyscf}.

\section{Acknowledgements}
The development of the MBGF-Net model was supported by the National Science Foundation under award number CHE-2337991 (C.V., T.Z.) and the National Science Foundation Engines
Development Award: Advancing Quantum Technologies (CT) under award number 2302908 (C.H.). The development of the Green's function property analysis and the BSE code was supported by the Air Force Office of Scientific Research under award number FA9550-24-1-0096 (J.L.). C.V. acknowledges partial support from the Department of Defense through the National Defense Science \& Engineering Graduate (NDSEG) Fellowship Program. J.L. acknowledges partial support from the Tony Massini Postdoctoral Fellowship in Data Science from Yale University. We thank the Yale Center for Research Computing for guidance and use of the research computing infrastructure.

\section{Author Contributions}
C.V. and T.Z. designed the project and wrote the manuscript. C.V. developed the graph neural network model and code. C.V., J.L., and T.Z. developed the Green's function post-processing workflow. J.L. and C.H. developed the BSE code. C.V., C.H., X.L.P., and J.Liu performed Green's function calculations and data analyses. T.Z. supervised the project. All authors contribute to the discussion of the results as well as the writing and editing of the manuscript.

\section{Competing Interests}
The authors declare no competing interests.


\FloatBarrier
\newpage

\bibliography{mlgf}

\end{document}